\documentstyle[aps,prl]{revtex}

\begin{document}
\preprint{HEP/123-qed}
\title{The Relationship between Extremum Statistics and Universal Fluctuations}  
\author{S. C. Chapman\footnote{sandrac@astro.warwick.ac.uk},
G. Rowlands}
\address{
Physics Dept. Univ. of Warwick,
Coventry CV4 7AL, UK
}
\author{
N. W. Watkins}
\address{British Antarctic Survey (NERC), Cambridge, CB3 0ET, UK}
\date{\today}
\maketitle
\begin{abstract}
The normalized probability density function (PDF) of global
measures of a large class of highly correlated systems has previously
been demonstrated to fall on a single non
Gaussian ``universal" curve. 
We  derive the functional form of the
``global" PDF
in terms of the ``source"
PDF of the individual events in the system.
A single parameter distinguishes the global PDF and is related to
the exponent of the source PDF.
When normalized, the global PDF is shown to be insensitive to this parameter
and importantly we obtain the previously demonstrated ``universality"
from an uncorrelated Gaussian source PDF.
The second and third moments of the
global PDF are more sensitive, providing
a powerful tool to probe the degree of complexity of
physical systems.
\end{abstract}

\narrowtext
\twocolumn
The study of systems exhibiting non Gaussian
statistics is of considerable current interest.
These statistics are observed to arise in finite sized  many body
systems exhibiting correlation over a broad range of scales. 
The apparent ubiquitous nature of this behavior has led to interest in
self organized criticality \cite{bakbook,jensenbook} as a paradigm; 
 other  highly correlated systems include fluid turbulence.
Two recent results have highlighted the connection between extremum statistics
and highly correlated systems. The probability density
function (PDF) of fluctuations in power needed
to drive an enclosed rotating turbulent fluid at constant angular frequency 
has been measured 
over 2 decades in Reynolds number. Intriguingly, when the PDF $P(E)$ of these
series of experiments were normalized
to the first two moments 
they were found to fall on a single non Gaussian
 ``universal" curve \cite{pinton1,pinton2}.
 This same universal curve was later identified in a study of
 the two dimensional
X-Y model, a numerical model for magnetization near the critical point
\cite{nature}. To obtain the universal curve, the PDF of a global measure,
namely the 
magnetization summed over the entire system, is again normalized to the
first two moments. It was suggested \cite{nature} that these two disparate systems
share the same statistics  as they are both critical. The 
functional form of the ``universal" curve was found for the X-Y model
and was shown to be of the form  
\cite{bramprl,ftippet,gumbook}
\begin{eqnarray}
P(E)=K (e^{y-e^y})^a\;\;\;\mbox{with}\;\;\;
y=b(E-s)\label{ft}
\end{eqnarray}
with $a=\pi/2$ and $K, b, s$  obtained by normalizing the curve to
the first two moments. Crucially, it was then demonstrated 
\cite{bramprl} that
this curve was also in reasonable agreement with appropriately chosen
normalized global
measures for a range of numerical models of 
highly correlated systems.
It was suggested that this behavior is 
related to the extremum statistics that
arises from a process that is highly
correlated. 

In this Letter we give a comprehensive analysis of extremum statistics
in the context of finite sized systems.
Our aim is to determine  the relationship between
the underlying ``source" PDF of a given process and the PDF of
some global measure. Given that events occur over
a range of sizes, and that each
event represents some quantity,
magnetization, or energy dissipation say, we obtain 
a relationship between the ``source" PDF of the event
size, and the PDF of a global measure,  the total magnetization,
or energy dissipation over the system.
We find, as suggested in \cite{bramprl}, that the
global PDF, when normalized to the first and second moments is essentially
of the form of equation (\ref{ft}). Crucially however we find that the
``universal" curve for the global PDF, that is, equation (\ref{ft})
with $a= \pi/2$ is not uniquely a property of a source PDF
of a correlated process. Instead, in a finite sized system,
distributions of this form with
$a$ in the range $[1,2]$ arise from uncorrelated samples from a source  
PDF 
ranging from exponential through Gaussian to power law, the value of $a$
being determined by  the source PDF.
When normalized to the first and second moments these curves are only
distinguishable asymptotically. 
Hence in reality the ``universal" curve describes, to within 
typical experimental
or numerical statistical uncertainties, distribution (\ref{ft})
with $a$ in the range $[1,2]$.

In many physical situations it is relatively straightforward
to measure the PDF of some global quantity such as power dissipation
in the driven turbulent fluid. In order to understand the underlying
process we require details of the distribution of the source PDF.
In particular, if this process is highly correlated, the source PDF of
individual events is anticipated to be power law and we
wish to i) distinguish this unambiguously from an uncorrelated Gaussian
process and ii) measure the exponent. A direct measurement of the source PDF
requires the challenging measurement of event sizes
over many decades, but if we can relate
the power law exponent to the form of the global PDF 
there is the
possibility to remote sense this exponent.  Normalizing the 
global PDF to the first and second moments is an
insensitive method to  find $a$;  we show that for finite sized systems
the higher order moments 
 provide a more feasible method.

The first step is to obtain the PDF of some global quantity from that of
the source PDF that describes individual events.
Consider a finite sized system 
of dimension $D$ which at any instant in time has patches of activity
on various length scales up to the system size $L_B$. The patches are drawn from the (time independent)
source probability $N(L)$
of a patch of length  $L$. These
patches can represent sites involved in
an avalanche in a sandpile, vortices in a turbulent fluid, ignited trees
in a forest fire, or sites with nonzero magnetization in the X- Y model.
Associated with the active sites is some quantity of interest, $Q$ say, 
for example energy
or magnetization, which we take to be given by $Q=L^D$. 
There will be some maximum $Q_B$ corresponding to the (extremely rare)
configuration with the highest possible value of $Q$, that is, highest energy
or magnetization, that can be realized by the system.
The total value of $Q$ over the system at any instant arises
from the distribution of the patches at that instant $N_j(L)$;
\begin{eqnarray}
\bar{Q}_j=\int_0^{Q_B} Q N_j(Q) dQ=\int_0^{L_B} L^D N_j(L) dL
\end{eqnarray}
where $N_j$ is the distribution of
an (unknown) realizable ensemble
of patches (continuous limit $N(L)$) that fits within the finite sized system, and
the integral is over the system. Since $N$ is  normalized,
$N(L)dL=N(Q)dQ$.
We now wish to evaluate the PDF of the $\bar{Q}_j$. This arises from the many
ensembles of the system, for the $j^{th}$  ensemble 
the total value of $Q$  can alternatively be written as
a sum over the $M_j$ (unknown) individual patches 
$\{L_{i}\}_j$,  $1\leq i\leq M_j$.
If $N_j(L)$ is monotonically decreasing
(from maximum $N_{0j}$ to zero)
we can generate each of the $\{L_{i}\}_j$ by choosing $M_j$ 
random numbers $N_i$ in the range $[0,N_{0j}]$, with uniform probability
distribution $P(N_i)$.
If we then insist that $P(N_i)\equiv P(L_i)$,
for each realization the random $N_i$ will each lie in one of the
$M_j$ uniform intervals
 $\delta N_i$, giving  $L_i$ patches which lie in  corresponding
(nonuniform)
intervals $\delta L_i$ obtainable in principle by inverting $N_j(L_i)$. 
We can then write the sum of the patches in the $j^{th}$ ensemble:
\begin{equation}
\bar{Q}_j=\sum_{i=1}^{M_j} \{L_{i}^D\}_j=\sum_{i=1}^{M_j}P(L_i)\delta L_i
L_i^D
\end{equation}
If the gradient of $N(L)$ is near monotonic,
 $\delta N_i/\delta L_i\simeq-(dN/dL)$ so that
\begin{equation}
\bar{Q}_j\simeq\sum_{i=1}^{M_j}\frac{P(N_i) \delta N_i L_i^D}{-
dN/dL}
\equiv \int_{N_{0j}}^1 \frac{L^D dN}{dN/dL}
\end{equation}
For a source PDF $N(L)$ that is
exponential, Gaussian or  inverse power law for large L  
 $dN/dL << L^D$ for small $N$, that is, large $L$ (large $Q$).
Hence the dominant contribution to  $\bar{Q}_j$
is that of the largest patch of activity. Thus the statistics of the
PDF of $\bar{Q}$, $P(\bar{Q})$ will be extremum statistics,
$P(\bar{Q})=P_m(Q)$,
 the normalized PDF of the maximum drawn from the
ensembles. 
Given that the maximum for the $j^{th}$ 
 ensemble is given 
by 
$Q^*_j=max\{Q_1,..Q_{M_j}\}$, where $Q_{M_j}\leq Q_B$, that is, $M_j$ finite,
 the PDF for $Q^{*}$ is given by
\begin{eqnarray}
P_m(Q^*)=MN(Q^*)(1-N_>(Q^*))^{M-1}
\end{eqnarray}
where $M$ is the average of $M_j$ over the ensembles and
\begin{eqnarray}
N_>(Q^*)=\int_{Q^*}^{Q_B} N(Q)dQ \simeq \int_{Q^*}^\infty N(Q)dQ
\end{eqnarray}
We  now obtain $P_m$ for large finite $M,Q$.
For a  general PDF $N(Q)$,
$(1-N_>)^{M}=\exp(-Mg(Q^*))$
where
\begin{equation}
g(Q^*)=-\ln(1-N_>(Q^*))\sim N_>+\frac{N_>^2}{2}\label{expN}
\end{equation}
We now  choose a characteristic value of $Q^*$,
namely $\tilde{Q}^*$, such that for any of the $j$ 
ensembles 
\begin{equation}
q=Mg(\tilde{Q}^*)=MN_>(\tilde{Q}^*)+M\frac{N_>^2(\tilde{Q}^{*})}{2}+\cdots\label{qdef}
\end{equation}
Using this definition and the form for $g(Q^*)$ (\ref{expN}) we obtain
$g'(\tilde{Q}^*)=-N(\tilde{Q}^*)$ to lowest order in an expansion in $q/M$.

We now consider specific source PDF $N(Q)$.
If $N(Q)$
falls off sufficiently fast in $Q$, i.e. is
Gaussian or exponential we can
consider lowest order only giving  $g(Q^*)\sim N_>$ \cite{gumbook,bouch}
 and $q=MN_>(\tilde{Q}^*)$. 
After some algebra, expanding in  
  $Q^*$ near $\tilde{Q}^*$ gives
\begin{equation}
P(\bar{Q})=P_m(Q) \equiv P_m(Q^*)\sim (e^{u-e^u})^a\label{gumgum}
\end{equation}
with
\begin{eqnarray}
a=\frac{N'(\tilde{Q}^*)N_>(\tilde{Q}^*)}{N^2(\tilde{Q}^*)}\\\label{agumb}
u=\ln(M N_>(\tilde{Q^*}))+\frac{N(\tilde{Q}^*)}{N_>(\tilde{Q}^*)}\Delta Q^*
\label{expu}
\end{eqnarray}
where $\Delta Q*=Q*-\tilde{Q^*}$.
For $N(Q)$ exponential (\ref{agumb}) gives $a\equiv 1$  (see \cite{bouch}).
For $N(Q)$ Gaussian we cannot obtain $a$ exactly 
but as we shall see it is instructive
to make an estimate. Given $N(Q)=N_0 \exp(-\lambda Q^2)$
in the above we obtain $P_m=\bar{P_m} \exp(R(u))$ with
\begin{eqnarray}
R=-\frac{\ln^2(q)}{4\lambda \tilde{Q}^{*2}}\label{Reqn}
+\bar{u}\left(1+\frac{2\ln (q)}{4\lambda \tilde{Q}^{*2}}\right)
-\frac{\bar{u}^2}{4\lambda \tilde{Q}^{*2}}
-e^{\bar{u}}
\end{eqnarray}
where we have used $u=-2\lambda \tilde{Q}^*\Delta Q^*$ and $\bar{u}=u+\ln(q)$.
To lowest order in $\Delta Q^*/\tilde{Q}^*$ (i.e. $\tilde{Q}^*\rightarrow \infty$)
we have  PDF (\ref{ft}) with $a=1$, but to next order, that is, neglecting 
the term in $\bar{u}^2$ only in (\ref{Reqn}) we have this PDF with
\begin{eqnarray}
a\equiv \left(1+\frac{2 \ln (q)}{4\lambda \tilde{Q}^{*2}}\right) \neq 1
\end{eqnarray}
Power law source PDF 
 $N(Q)$ fall off sufficiently slowly with $Q$ that we need
to go to next order in $\Delta Q^*/\tilde{Q^*}$.
 If we consider normalizable source PDF
\begin{equation}
N(Q)=\frac{N_0}{(1+Q^2)^k}\label{power1}
\end{equation}
then for large $Q$ the above method yields that
$P(\bar{Q})$ is given by the form  (\ref{gumgum}) but with
\begin{equation}
u=-\ln(a)-\ln(q)-(2k-1)\frac{\Delta Q^*}
{\tilde{Q}^*}(1-\frac{\Delta Q^*}{2\tilde{Q}^*})\label{baru}
\end{equation}
and 
$a=2k/2k-1$.
To lowest order, neglecting the $(\Delta Q^* / \tilde{Q}^*)^2$
term 
 (\ref{baru}) reduces to 
(\ref{expu}).
 Hence  a power law source PDF
has maximal statistics
$P_m(Q)$ which, when evaluated to next order, 
have distribution (\ref{ft}) with a correction that is non negligible
  at the asymptotes,  consistent with
  the well known result due to Frechet (\cite{bouchbook,jenkinson}).

The above results should be contrasted with that of Fischer and Tippett
\cite{ftippet}.
Central to  \cite{ftippet} and later derivations is that 
a single ensemble of $NM$ patches has the same statistics as
the 
 $N$ ensembles 
(of $M$ patches), of which it is comprised.  
The fixed point of this expression for arbitrarily large $N$ and $M$ 
is $a=1$ for the exponential
and Gaussian PDF, and the Frechet result for power law PDF.
Here, we consider a finite sized system so that although the number
of realizable ensembles of the system can be taken arbitrarily large, the
number of patches $M$ per ensemble is always large but finite.
Importantly, the rate of convergence with $M$ depends on the
PDF $N(L)$. For an exponential or power law PDF
we are able to resum the above expansion exactly to obtain
$a$; and convergence will then just depend on terms $O(1/M)$ and above.
This procedure is not possible for $N(Q)$ Gaussian, instead we 
consider the characteristic $Q^*$, that is $\tilde{Q}^*$ 
which for  $M$ arbitrarily large should be large also.
Rearranging
(\ref{qdef}) to lowest order for  $N(Q)=N_0 \exp(-\lambda Q^2)$
yields $\sqrt \lambda \tilde{Q}^* \sim
\sqrt{ln(M)}$ implying significantly slower convergence.

We now have the intriguing result that
for a wide range of source PDF the
PDF of a global measure $P(\bar{Q})$ is essentially a family
of curves that are approximately
Gumbel in form and are asymmetric with a handedness
 that just depends on the sign of $Q$; we have assumed $Q$ positive
 whereas one could choose $Q$ negative (with $L$ positive) in which case
 $N(Q)\rightarrow N(\mid Q \mid)$. 
The single parameter $a$ that distinguishes
the global PDF then just depends on the source PDF of
the individual events. For $N(Q)$ exponential we recover the
well known result  \cite{gumbook,bouch} $a=1$.  
For a power law source PDF $a$ is determined by $k$ as above.
For a Gaussian source PDF $a \neq 1$.

To compare these curves we normalize
$P(\bar{Q})\equiv P_m(Q^*)$. 
For Gaussian and exponential source PDF we have
\begin{eqnarray}
\bar{P}(y)=K  (e^{u-e^u})^a\;\;\;\;\mbox{with}\;\;\;\;
u=b(y-s)\label{expu2}
\end{eqnarray}
This has moments
\begin{equation}
M_n=\int_{-\infty}^\infty y^n \bar{P}(y) dy\label{moment}
\end{equation}
which converge for all $n$; 
we insist that $M_0=1$,  $M_1=0$ and $M_2=1$.
The necessary integrals can be expressed in terms of derivatives
of the Gamma function $\Gamma(a)$ \cite{gamref}
and
we obtain after some algebra 
\begin{eqnarray}
b^2= \Psi'(a),\;\;\;
K=\frac{b}{\Gamma(a)}e^{a \ln(a)},\;\;\;
s= - \frac{1}{b}(\Psi(a)-\ln(a))\label{gumnorm}
\end{eqnarray}
where
$\Psi(a)$ and its derivative w.r.t. $a$,
$\Psi'(a)$ have their usual meaning.
The ambiguity in the sign of $b$ (and hence $s$) corresponds to
the two solutions for $P(\bar{Q})$ for positive and negative $Q$.

For power law source PDF (\ref{power1}) we use
the Frechet distribution (\cite{bouchbook,jenkinson}) which we first write as (\ref{expu2}) with
\begin{equation}
u=\alpha+\beta \ln(1+\frac{y}{G})\label{frechu}
\end{equation}
which reduces to the form of (\ref{expu2}) for 
$\Delta Q^*/\tilde{Q}^* \ll 1$. 
From (\ref{power1},\ref{baru}) we identify
$\beta=-(2k-1)$.
Again we insist that $M_0=1$,  $M_1=0$ and $M_2=1$
and obtain
\begin{eqnarray}\nonumber
\alpha=-\beta \ln\left(\frac{a^{\frac{1}{\beta}}}{\Gamma(1+1/\beta)}\right)\\
K=\pm \beta a^a \left[\Gamma(1+2/\beta)-\Gamma^2(1+1/\beta)
\right]^\frac{1}{2}\label{frechnorm}\\
G=\frac{\Gamma(1+\frac{1}{\beta})}{\left[\Gamma(1+\frac{2}{\beta})-\Gamma^2(1+\frac{1}{\beta})
\right]^\frac{1}{2}}\nonumber
\end{eqnarray}
For $\beta\rightarrow\infty$ with $\beta/G$ finite these equations reduce to
(\ref{gumnorm}) with $b=-\beta/G$.

We can now plot the ``universal" curves, that is, normalized to the
first two moments. Experimental measurements of
a global PDF $P(E)$ normalized to $M_0$
would be plotted $M_2 P$ versus $(E-M_1)/M_2$.
For the Frechet
it is straightforward to show that the moments of order $n$ (\ref{moment})
exist for $2k>n+1$ and therefore these curves 
exist for power law of index $\infty > 2k > 3$ i.e.
$ 1 < a < 3/2$. 
This is
 significant since processes exhibiting 
 long range correlations typically have $k$ lower than this \cite{jensenbook}.
Inset in Figure 1 we plot the normalized Frechet PDF for $k=3,5,100$  
and the  PDF (\ref{ft}) with $a=1$. 
In the  limit $k\rightarrow\infty$, $a\rightarrow 1$ and the normalized
Frechet PDF tends to the $a=1$ limit of (\ref{ft}),
hence for $k=100$ these are indistinguishable
and differences between the  PDF
 appear on such a plot around the mean for $k<3$ approximately.
 In the main plot we show normalized distributions of the form (\ref{ft})
  for $a=1,\pi/2$ and
  $2$. It is immediately apparent that the
  curves are difficult to distinguish for several decades
  in $\bar{P}(y)$ and  either numerical or real experiments would require
  good statistics over a dynamic range of about 4 decades which is not
  readily achievable.

Since the second moment $M_2$ does not exist for $k\leq 3/2$  we cannot
consider curves of $a \geq 3/2$ generated by power law source PDF;
however  such values (in particular $a=\pi/2$) were identified for the
``universal" curves in turbulence
experiments and a variety of models of correlated systems
\cite{nature,bramprl}. We now demonstrate
that these are straightforward to produce.
 On Figure 1 we have over plotted (*) the global PDF generated by 
a source PDF that is uncorrelated Gaussian, calculated
 numerically. We randomly select $M$ uncorrelated variables
 $Q_{j},j=1,M$ and to specify the handedness of the extremum distribution,
 the $Q_j$ are defined negative and $N(\mid Q \mid)$
 is normally distributed. This would physically
  correspond to a system where the global quantity $\bar{Q}$ is negative,
   i.e. power consumption in a turbulent fluid, as opposed to power generation.
   To construct the global PDF we generate  $T$ ensembles, that is select
 $T$ samples of the largest negative number $Q^*_i=min\{Q_1..Q_M\}, i=1,T$.
 For the data shown in the figure $M=10^5$ and $T=10^6$; this gives
 $\sqrt \lambda \tilde{Q}^* \sim \sqrt{ln(M)}\simeq 3$ so that for the
 Gaussian we are far from the $a=1$ limit \cite{ftippet}.

 The numerically calculated PDF lies close to $a=\pi/2$.
 Such a value of $a$ on these ``universal" curves is therefore not
 strong evidence of a {\em correlated} process as suggested by \cite{bramprl}.
Generally, plotting data in this way is an 
insensitive method for determining $a$ 
and thus distinguishing the statistics of the underlying physical process.

The question of interest is whether we can determine the form
of the source PDF from the global PDF
from data with
 a reasonable dynamic range. 
 We consider two possibilities here.
First, a uniformly sampled process will have the most statistically significant
values on the universal curve near the peak. 
For both the  PDF the peak is at $u=0$ 
and is at $\bar{P}(u=0)=Ke^{-a}$ with $K$ given by (\ref{gumnorm})
and (\ref{frechnorm}) respectively.
The latter applies to $k>3/2$; for smaller $k>1$ we may use $M_0=1,M_1=0$ 
plus a condition on $\bar{P}(u=0)$ to obtain $a$.
A more sensitive indicator
may be the third moment of $\bar{P}$  
 which after some algebra
can be written as
\begin{eqnarray}
M_3=-\frac{\Psi''(a)}{(\Psi'(a))^\frac{3}{2}}
\end{eqnarray}
for a Gaussian or exponential source PDF i.e. with (\ref{expu2})
and
\begin{eqnarray}
M_3=\frac{\left[\Gamma(1+\frac{3}{\beta})-3\Gamma(1+\frac{2}{\beta})
\Gamma(1+\frac{1}{\beta})+2\Gamma^3(1+\frac{1}{\beta})\right]}
{\left[\Gamma(1+\frac{2}{\beta})-\Gamma^2(1+\frac{1}{\beta})\right]^\frac{3}{2}}
\end{eqnarray}
for a power law source PDF i.e. with (\ref{frechu}); the latter  converging
for $2k>4$.
Again these refer to one of the two possible solutions for $P(\bar{Q})$;
the other solution 
 corresponding to $y\rightarrow -y$,
$M_3 \rightarrow -M_3$. 
 We can compare these
two methods by noting that for  PDF of the form (\ref{ft}) with
$a=1,2$ the corresponding values of
$\bar{P}_m$  differ by $\sim 7.9$\% whereas $M_3$ differs by
$\sim 32$\%. 
For Frechet PDF, the variation in $\bar{P}(u=0)$ is most significant
for smaller $k$, for example with $k=3,4$ $\bar{P}(u=0)$ differs by $\sim 15\%$
whereas $M_3$ differs by $\sim 30 \%$.

In conclusion, we have shown that the statistics of fluctuations in
a global measure of a finite sized system, such as total energy dissipation  in a turbulent fluid,
or total magnetization in a ferromagnet
are generally given by extremum statistics.
The PDF
of the global measure is then one of a 
family of curves whose moments have been determined in terms of a single
parameter $a$ which in turn quantifies the PDF of the underlying 
``source" process, such as the PDF of individual energy release events
or patches of magnetization.
When normalized to the first and second moments
these curves are insensitive to $a$ and 
fall close to the  single ``universal" curve previously
identified as a property of a large class of highly correlated systems \cite{bramprl}, over
the range achieved by previous real or numerical experiments.
In particular, we find that the global PDF of an uncorrelated Gaussian 
process is 'Gumbel' (\ref{ft}) distributed with $a \simeq \pi/2$, providing
a straightforward explanation for the previously demonstrated ``universality".
Finally we suggest that the peak, or 
the third moment of the
global PDF is
a more sensitive indicator of the source PDF. This is
a powerful tool to probe the exponents of
physical systems where the source PDF is difficult to
measure but provides a signature of the degree of complexity of the system.

\acknowledgments
The authors  thank G. King and M. Freeman
for illuminating discussions.
SCC was supported by PPARC.

\begin{figure}
\caption{The normalized  PDF (\ref{ft}) with $a=2,\pi/2,1$ (the
right hand asymptotes of the curves intersect the ordinate in that order from
left to right).
Overplotted
is the numerically evaluated global PDF of an uncorrelated Gaussian process.
 Inset are Frechet PDF normalized to the first two moments for
 source PDF exponents $2k$,  $k=3,5,100$, on the same scale.}
\label{autonum}
\end{figure}

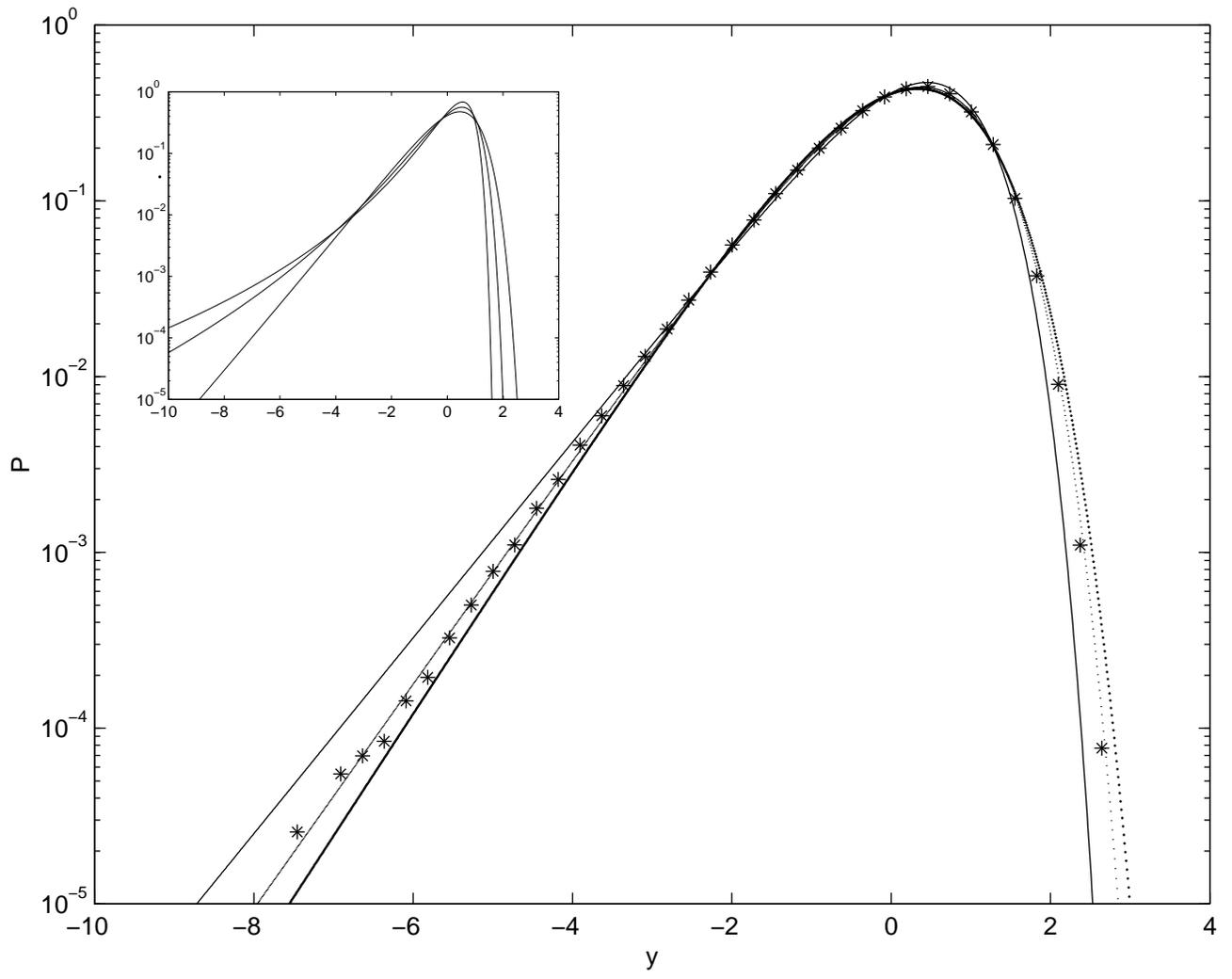
\end{document}